\documentclass[a4paper,fleqn,usenatbib,useAMS]{mnras}
 
\usepackage{mathptmx}
\usepackage{txfonts}

\usepackage[T1]{fontenc}
\usepackage{ae,aecompl}

\usepackage{graphicx}	
\usepackage{amssymb}	
\usepackage{epsfig}

\usepackage{natbib}
\usepackage{footnote}
\usepackage{lscape}
\usepackage{longtable}
\usepackage{supertabular}
\usepackage{caption}
\usepackage{subfig}
\usepackage{txfonts}

\title[Modeling of the outburst on July 29th, 2015.]{Modeling of the outburst on July 29th, 2015 observed with OSIRIS cameras in the southern hemisphere of comet 67P/Churyumov-Gerasimenko.}

\author[A. Gicquel et al.]{
\parbox{\textwidth}{
\begin{normalsize}
A. Gicquel$^{1,2}$\thanks{E-mail: adeline.gicquel@jpl.nasa.gov}, M. Rose$^{3}$,
J.-B. Vincent$^{4}$, B. Davidsson$^{2}$, D. Bodewits$^{5}$, M. F. A'Hearn$^{5}$, J. Agarwal$^{1}$, N. Fougere$^{6}$, H. Sierks$^{1}$, I. Bertini$^{7}$,  Z.-Y. Lin$^{8}$, C. Barbieri$^{9}$, P. L. Lamy$^{10}$, R. Rodrigo$^{11,12}$, D. Koschny$^{13}$, H. Rickman$^{14,15}$, H. U. Keller$^{16}$, M. A. Barucci$^{17}$, J.-L. Bertaux$^{18}$, S. Besse$^{13}$, S. Boudreault$^{1}$, G. Cremonese$^{19}$, V. Da Deppo$^{20}$, S. Debei$^{21}$, J. Deller$^{1}$, M. De Cecco$^{22}$, E. Frattin$^{19}$, M. R. El-Maarry$^{23}$, S. Fornasier$^{17}$, M. Fulle$^{24}$, O. Groussin$^{10}$, P. J. Guti\'errez$^{25}$, P. Guti\'errez-Marquez$^{1}$, C. G\"uttler$^{1}$, S. H\"ofner$^{1,16}$, M. Hofmann$^{1}$, X. Hu$^{1}$, S. F. Hviid$^{4}$, W.-H. Ip$^{8}$, L. Jorda$^{10}$, J. Knollenberg$^{4}$, G. Kovacs$^{1,27}$, J.-R. Kramm$^{1}$, E. K\"uhrt$^{4}$, M. K\"uppers$^{27}$, L. M. Lara$^{25}$, M. Lazzarin$^{9}$, J. J. Lopez Moreno$^{25}$, S. Lowry$^{28}$, F. Marzari$^{9}$, N. Masoumzadeh$^{1}$, M. Massironi$^{7}$, F. Moreno$^{25}$, S. Mottola$^{4}$, G. Naletto$^{29,7,22}$, N. Oklay$^{4}$, M. Pajola$^{30}$, A. Pommerol$^{23}$, F. Preusker$^{4}$, F. Scholten$^{4}$, X. Shi$^{1}$, N. Thomas$^{23}$, I. Toth$^{31,10}$, C. Tubiana$^{1}$
\end{normalsize}} 
\\~\\
\parbox{\textwidth}{
$^{1}$Max-Planck Instit\"ut fur Sonnensystemforschung, Justus-von-Liebig-Weg 3, 37077 G\"ottingen, Germany; $^{2}$Jet Propulsion Laboratory/California Institute of Technology, 4800 Oak Grove Dr., Pasadena, California 91109, USA;  $^{3}$Ingenieurbuero Dr.-Ing. Martin Rose, Goldmuehlestr. 6, 71065 Sindelfingen, Germany;  $^{4}$Institute of Planetary Research, DLR, Rutherfordstrasse 2, 12489 Berlin, Germany; $^{5}$Department for Astronomy, University of Maryland, College Park, MD 20742-2421, USA;  $^{6}$Department of Climate and Space Sciences and Engineering, University of Michigan, Ann Arbor, MI 48109, USA; $^{7}$Centro di Ateneo di Studi ed Attivit\'{a} Spaziali "Giuseppe Colombo" (CISAS), University of Padova, Via Venezia 15, 35131 Padova, Italy; $^{8 }$Institute for Space Science, National Central University, 32054 Chung-Li, Taiwan; $^{9}$Department of Physics and Astronomy "G. Galilei", University of Padova, Vic. Osservatorio 3, 35122 Padova, Italy; $^{10}$Aix Marseille Universit\'e, CNRS, LAM (Laboratoire d'Astro-physique de Marseille) UMR 7326, 13388, Marseille, France; $^{11}$Centro de Astrobiologia (INTA-CSIC), European Space Agency, European Space Astronomy Centre (ESAC), P.O. Box 78, E-28691 Villanueva de la Canada, Madrid, Spain; $^{12}$International Space Science Institute, Hallerstrasse 6, 3012 Bern, Switzerland; $^{13}$Research and Scientific Support Department, European Space Agency, 2201 Noordwijk, The Netherlands; $^{14}$Department of Physics and Astronomy, Uppsala University, Box 516, 75120 Uppsala, Sweden; $^{15}$PAS Space Research Center, Bartycka 18A, 00716 Warszawa, Poland;  $^{16}$Institute for Geophysics and Extraterrestrial Physics, TU Braunschweig, 38106 Braunschweig, Germany; $^{17}$LESIA, Observatoire de Paris, CNRS, UPMC Univ Paris 06, Univ. Paris-Diderot, 5 Place J. Janssen, 92195 Meudon Pricipal Cedex, France; $^{18}$LATMOS, CNRS/UVSQ/IPSL, 11 Boulevard d'Alembert, 78280 Guyancourt, France; $^{19}$INAF Osservatorio Astronomico di Padova, Vicolo dell'Osservatorio 5, 35122 Padova, Italy; $^{20}$CNR-IFN UOS Padova LUXOR, Via Trasea 7, 35131 Padova, Italy; $^{21}$Department of Industrial Engineering University of Padova Via Venezia, 1, 35131 Padova, Italy; $^{22}$University of Trento, via Sommarive, 9, Trento, Italy; $^{23}$Physikalisches Institut, Sidlerstrasse 5, University of Bern, CH-3012 Bern, Switzerland; $^{24}$INAF - Osservatorio Astronomico di Trieste, via Tiepolo 11, 34143 Trieste, Italy; $^{25}$Instituto de Astrofisica de Andalucia-CSIC, Glorieta de la Astronomia, 18008 Granada, Spain; $^{26}$Budapest University of Technology and Economics, Department of Mechatronics, Optics and Engineering Informatics, Muegyetem rkp 3, Budapest, Hungary; $^{27}$ESA/ESAC, PO Box 78, 28691 Villanueva de la Ca\~nada, Spain; $^{28 }$Centre for Astrophysics and Planetary Science, School of Physical Sciences, The University of Kent, Canterbury CT2 7NH, United Kingdom; $^{29}$Department of Information Engineering, University of Padova, Via Gradenigo 6/B, 35131 Padova, Italy; $^{30}$NASA Ames Research Center, Moffett Field, CA 94035, USA; $^{31}$Observatory of the Hungarian Academy of Sciences, PO Box 67, 1525 Budapest, Hungary.
}}

\date{Accepted XXX. Received YYY; in original form ZZZ}

\pubyear{2016}

\begin{document}
\label{firstpage}
\pagerange{\pageref{firstpage}--\pageref{lastpage}}
\maketitle

\begin{abstract}
Images of the nucleus and the coma (gas and dust) of comet 67P/Churyumov- Gerasimenko have been acquired by the OSIRIS (Optical, Spectroscopic, and Infrared Remote Imaging System) cameras since March 2014 using both the Wide Angle Camera (WAC) and the Narrow Angle Camera (NAC). We use images from the NAC camera to study a bright outburst observed  in the southern hemisphere on July 29, 2015.  The high spatial resolution of the NAC is needed to localize the source point of the outburst on the surface of the nucleus.  The heliocentric distance is 1.25 au and the spacecraft-comet distance is 186 km. Aiming to better understand the physics that led to the outgassing, we used the Direct Simulation Monte Carlo (DSMC) method to study the gas flow close to the nucleus and the dust trajectories. The goal is to understand the mechanisms producing the outburst. We reproduce the opening angle of the outburst in the model and constrain the outgassing ratio between the outburst source and the local region. The outburst is in fact a combination of both gas and dust, in which the active surface is approximately 10 times more active than the average rate found in the surrounding areas. We need a number of dust particles 7.83 $\times$ $10^{11}$ -  6.90 $\times$ $10^{15}$ (radius 1.97 - 185 $\mu$m), which corresponds to a mass of dust 220 - 21 $\times$ $10^{3}$kg. 
\end{abstract}

\begin{keywords}
comets: individual:67P/Churyumov-Gerasimenko -- methods: data analysis -- methods: observational -- methods: numerical
\end{keywords}



\section{Introduction}
The ESA (European Space Agency) Rosetta spacecraft was launched on March 2, 2004 and reached comet 67P/Churyumov-Gerasimenko (67P) in August of 2014. Since then, images of the nucleus and the coma  have been acquired by the OSIRIS (Optical, Spectroscopic, and Infrared Remote Imaging System) camera system \citep{Keller_2007} using both the wide angle camera (WAC) and the narrow angle camera (NAC). Close to perihelion in August 2015, a display of outbursts on 67P, known as the summer fireworks, was observed \citep{Vincent_2016}. The ESA's Rosetta spacecraft had the unique opportunity to follow the activity and morphology of comet 67P during its journey toward the Sun. \\

Many studies have presented the activity of the nucleus, such as localized dust and gas jets \citep{Lara_2015, Lin_2015, Lin_2016, Gicquel_2016}. During the three months surrounding the comet's perihelion passage in August 2015, \cite{Vincent_2016} reported the detection of 34 outbursts with one on average every 2.4 nucleus rotations (30 hours). On February 19, 2016, an outburst of gas and dust was monitored simultaneously by instruments onboard Rosetta and ground-based telescopes \citep{Grun_2016}. On July 3, 2016, another outburst was observed by many instruments onboard Rosetta \citep{Agarwal_2017}. \cite{Vincent_2016} defined an outburst as a bright event having a very short duration with respect to the rotation period of the nucleus. The increase of the brightness of the coma is due to the release of gas and dust, and it is typically one order of magnitude brighter than the usual jets. Also, due to the short lifetime, the outburst might be observable in one image only, depending on the observing cadence. \\

The present work analyzed if the opening angle of an outburst observed with the OSIRIS data could be reproduced using a Direct Simulation Monte Carlo (DSMC) method. We analyzed the outburst observed in the southern hemisphere of comet 67P on July 29, 2015 with the NAC two weeks before perihelion on August 13, 2015.  We studied the brightness distribution of the outburst (B [W $m^{-2}$ $nm^{-1}$ $sr^{-1}$ ]) as a function of the distance from the nucleus (D [km]). We presented the observations obtained with the OSIRIS cameras and described the method used to  reproduce the opening angle of the outburst, by first simulating just the gas (water) and then adding the dust. Finally, we compared the NAC image with the synthetic images. \\

.

\section{Observations}
\begin{figure*}
  \centering
  \subfloat[2015-07-29T12:49:28]{\label{NACa}\includegraphics[scale=0.2]{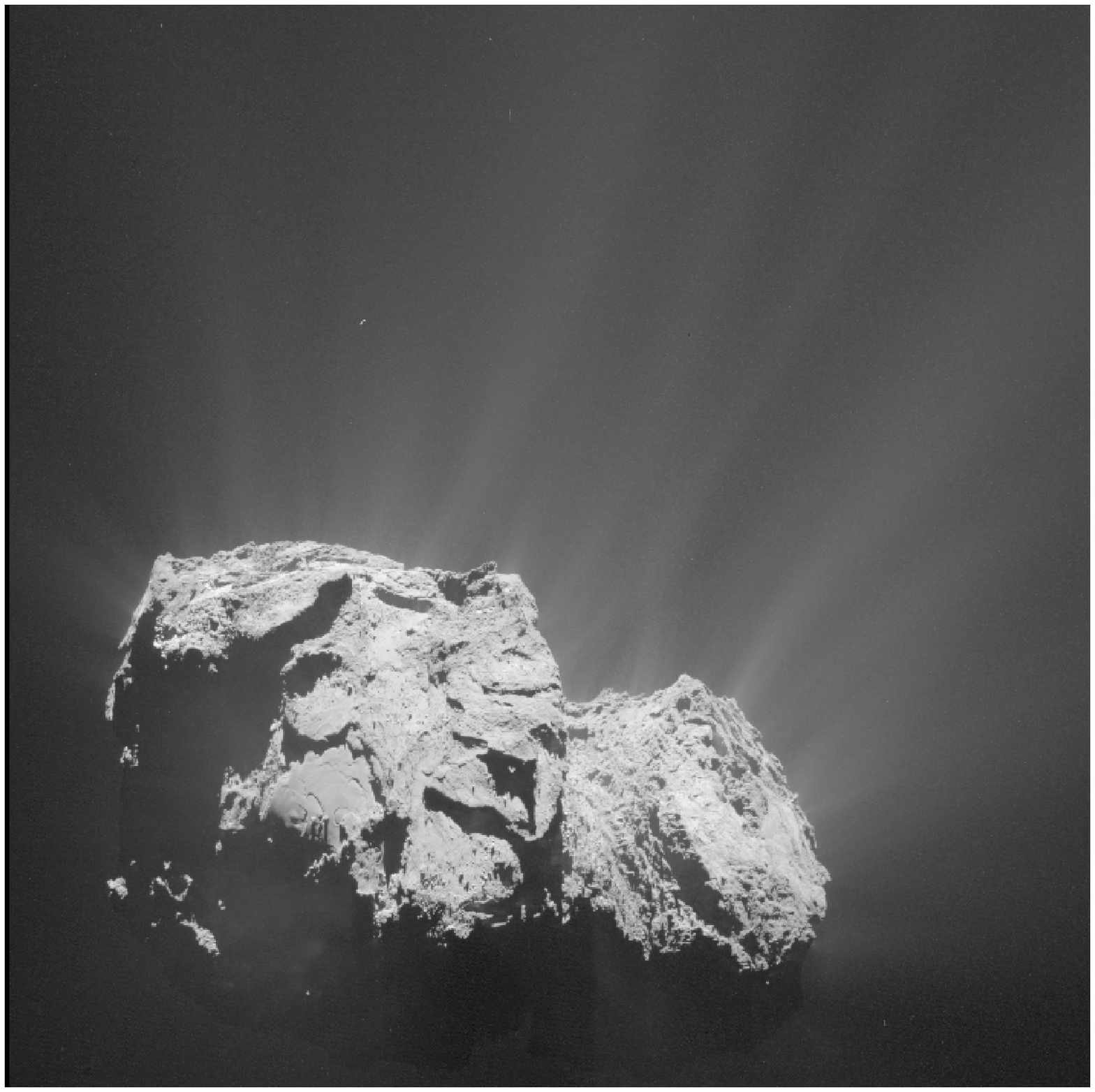}} 
  \subfloat[2015-07-29T13:07:28]{\label{NACb}\includegraphics[scale=0.2]{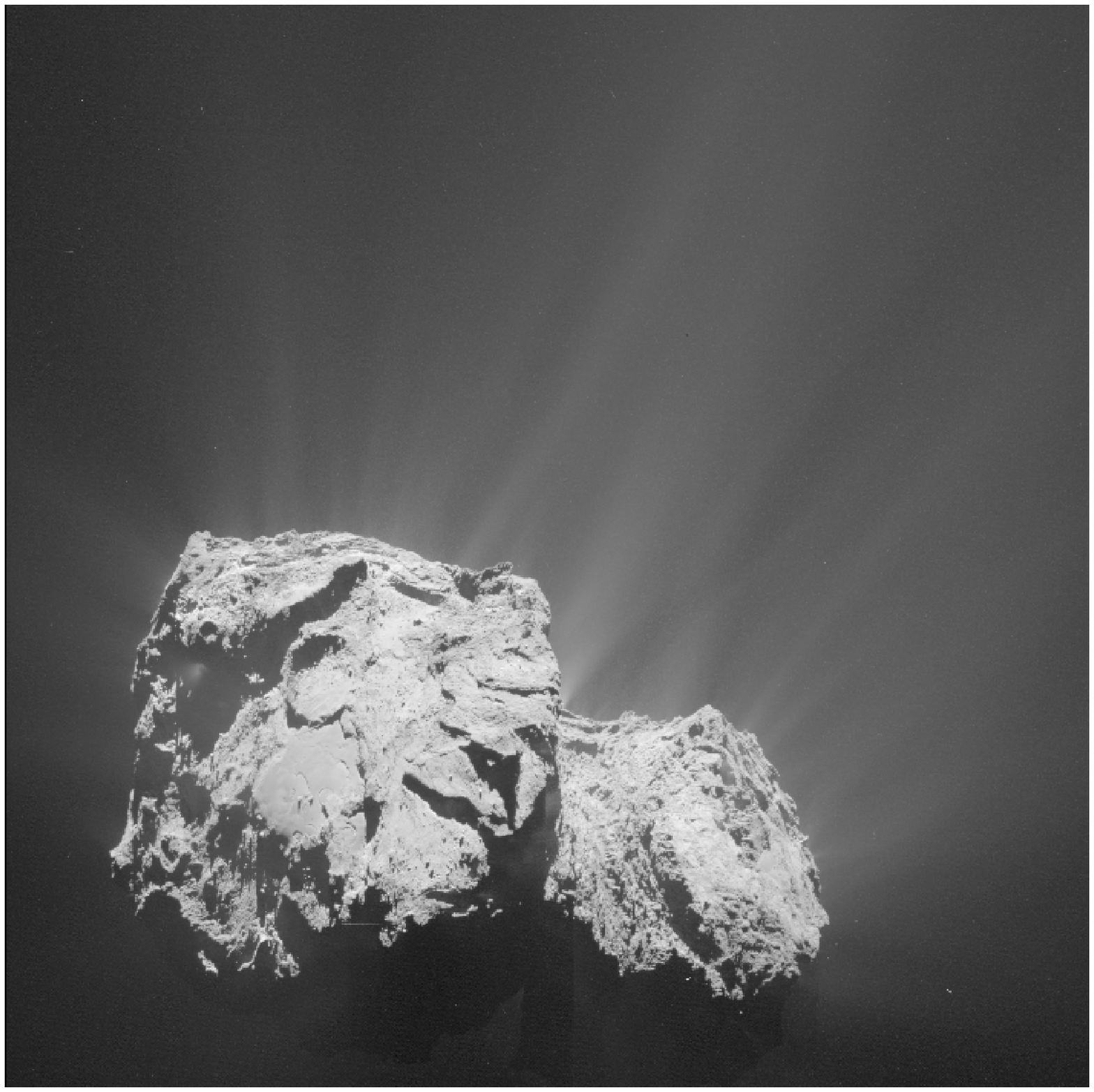}} 
\subfloat[2015-07-29T13:25:28]{\label{NACc}\includegraphics[scale=0.174]{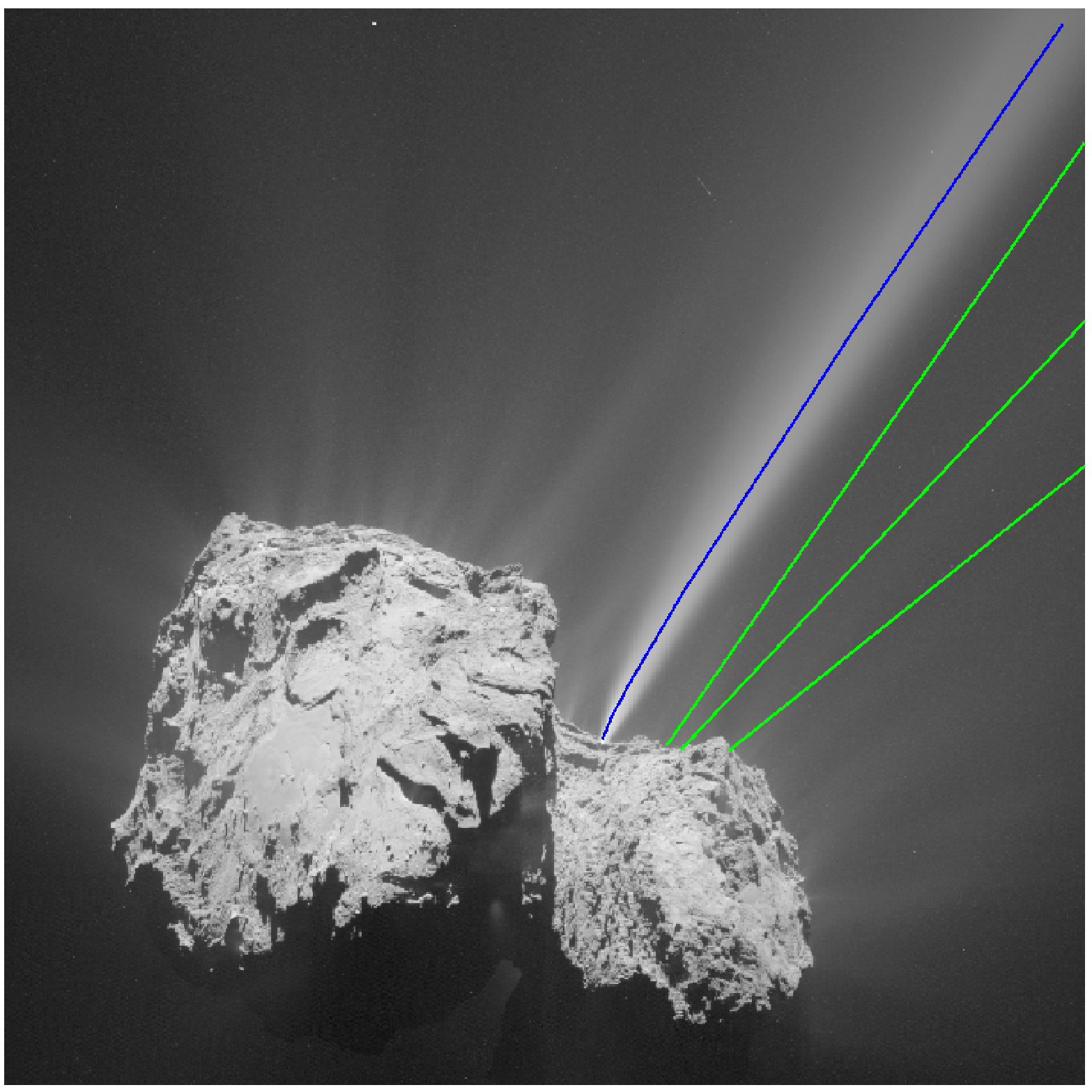}}
	\subfloat[2015-07-29T13:43:28]{\label{NACd}\includegraphics[scale=0.2]{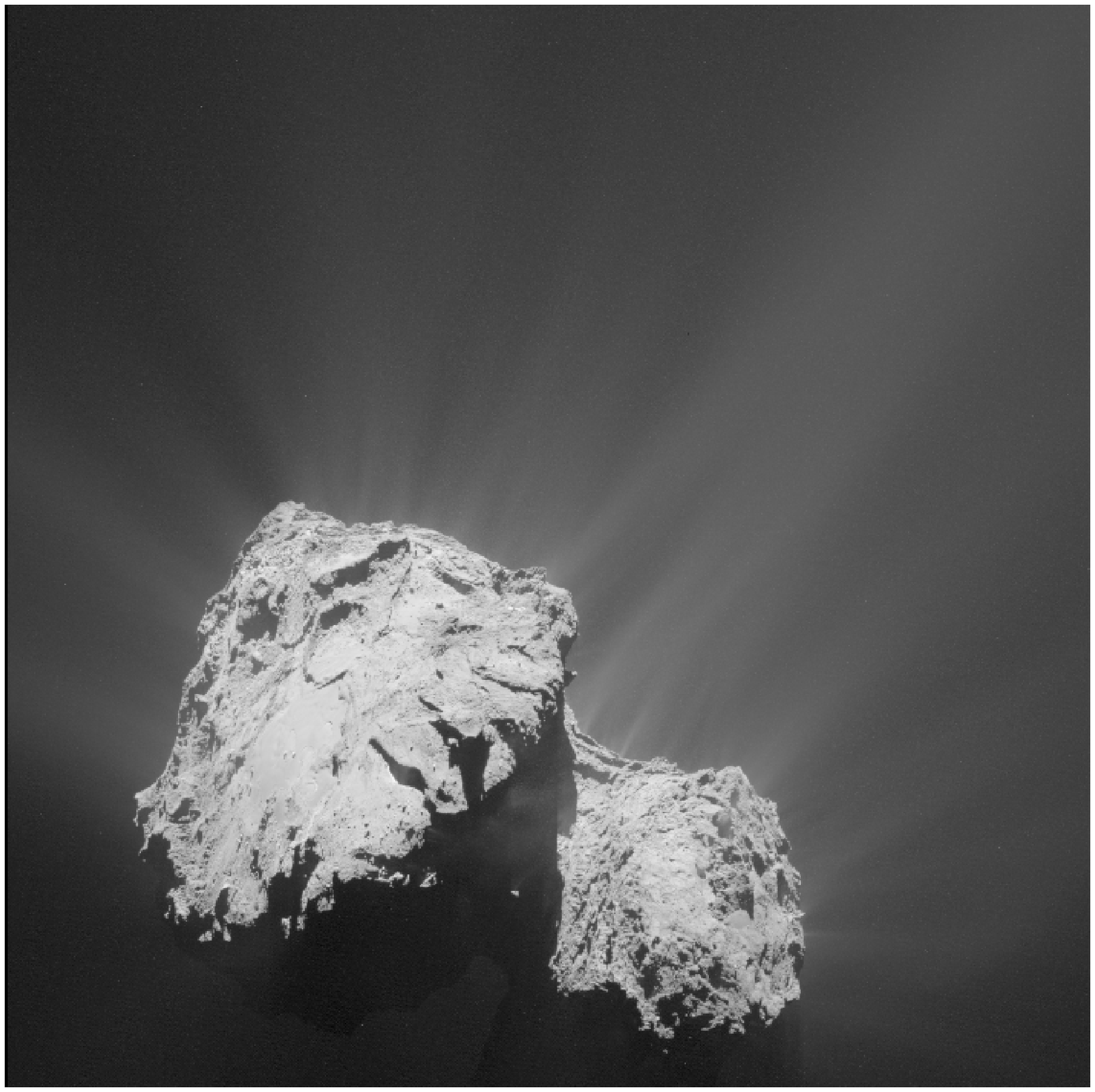}} 
	\subfloat[2015-07-29T14:01:28]{\label{NACe}\includegraphics[scale=0.2]{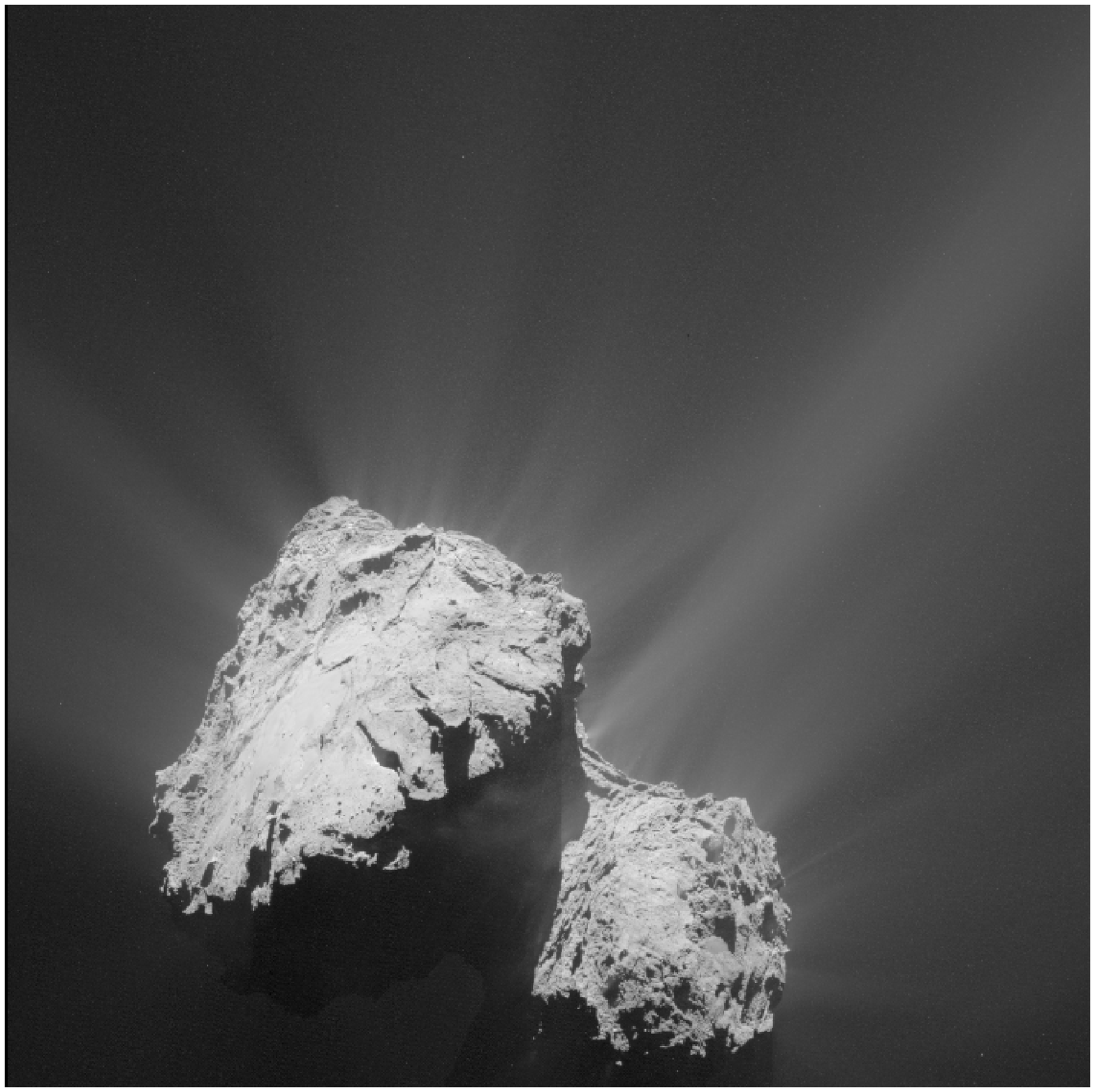}} 
\caption{The OSIRIS NAC images, the radial profile for the jet (blue) and the radial profile for the background coma (green)}
  \label{Observations_NAC}
\end{figure*}

\subsection{Data with the OSIRIS Cameras}
The OSIRIS cameras, composed of the WAC and NAC, were dedicated to mapping the nucleus of comet 67P and to characterizing the evolution of the comet's gas and dust \citep{Keller_2007}. The WAC (230 - 750 nm) was mainly used to study the coma of dust and gas, while the NAC (250 - 1000 nm) was used to investigate the structure of the nucleus. \\

We chose the monitoring observations on UT 13:25:28 July 29, 2015  utilizing the NAC orange filter (F22, center wavelength = $\lambda$ = 649.2nm, FWHM = 84.5nm). At the end of July and in August, the time line was densely covered with observations, and the gaps in outburst detection could not be explained by a lack of imaging. As shown in Figure \ref{Observations_NAC}, with a cadence imaging around 16 min, the outburst was detectable in Figure \ref{NACc} but not in Figures \ref{NACa}, \ref{NACb}, \ref{NACd} and \ref{NACe}. This bright outburst was emerging from the side of the comet's neck, in the Sobek region between two hills \citep[Figure 7b;][]{Vincent_2016}. We refer the reader to \cite{Thomas_2015} and \cite{ElMarry_201} for the nucleus map which indicates the regions. The outburst was observed 3.69 hours after sunrise (around local mid-day).\\

The outburst is classified as a Type A by \cite{Vincent_2016}, having a very collimated outburst where the dust and gas are ejected at high velocity. The high spatial resolution is needed to localize the source point of the outburst on the surface of the nucleus. The source location of the outburst, latitude = -37$^{\circ}$ and longitude = 300$^{\circ}$, is given by \cite{Vincent_2016} in the standard 
"Cheops" frame \citep{Preusker_2015}. The outburst probably originates from a small and confined area. The heliocentric distance is $R_h$ = 1.256 au, the spacecraft-comet distance is $\Delta_{S/C}$ =186 km and the resolution is 1.87 $\times$ 10$^{-5}$ rad pixel$^{-1}$. The pixel scale is 3.42 m px$^{-1}$ and the NAC field of view is (FOV) = 7 $\times$ 7 km. No binning was used in collecting or downlinking the images. Only one other outburst, no. 34, was observed approximately two months later, by the NAVCAM in the Sobek region on 2015-09-26T12:03:32 at latitude = -40$^{\circ}$ and  longitude = +307$^{\circ}$ \citep{Vincent_2016}.\\

\begin{figure}
  \centering
  \includegraphics[scale=0.17]{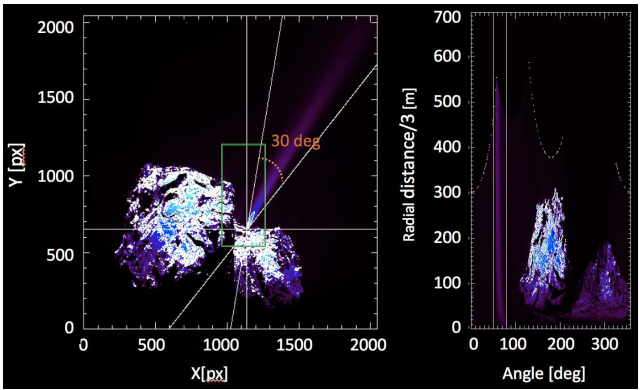}
    \vspace{0.1in}
    \caption{Size of NAC images (px), opening angle (30 deg) and length ($\approx$ 2.5 km) of the outburst on 2015-07-29T13:25:28 in Cartesians (left) and Polar coordinates (right). The green  box represents the size (315 x 585 pixels) and position of the synthetics images.}
    \label{VIRTIS_Model} 
\end{figure}

As shown in Figure \ref{VIRTIS_Model}, the size of the NAC image observed with the NAC camera on July 29th, 2015 is 2048 x 2048 pixels. In order to constrain the opening angle of the outburst, we switched from a Cartesian to Polar coordinate system. In Figure \ref{VIRTIS_Model}. the Cartesian coordinates are on the left side and the polar coordinate are on the right.  On the left side of the figure, where we used Cartesian coordinates, there are two white lines with an opening angle of 30 degrees.  You can see that the opening angle and the whole of the outburst are within these two lines. This corresponds to the vertical white line on the right side of Figure \ref{VIRTIS_Model}, where polar coordinates were used.  In both cases, you can see that the outburst is collimated.  \\

\subsection{Radial profiles}

In the present section, we aim to study the brightness distribution of the outburst as a function of distance from the nucleus. As explained by \cite{Gicquel_2016}, we average 3 radial profiles of the background  coma in the same area as the outburst, as shown in Figure \ref{NACc} (in blue). The radial profile is taken from the individual pixels along the center-line of the outburst, as shown in Figure \ref{NACc} (in green). The coma background is subtracted from the radial profile of the outburst. \\

\begin{figure}
\centering
\includegraphics[scale=0.29]{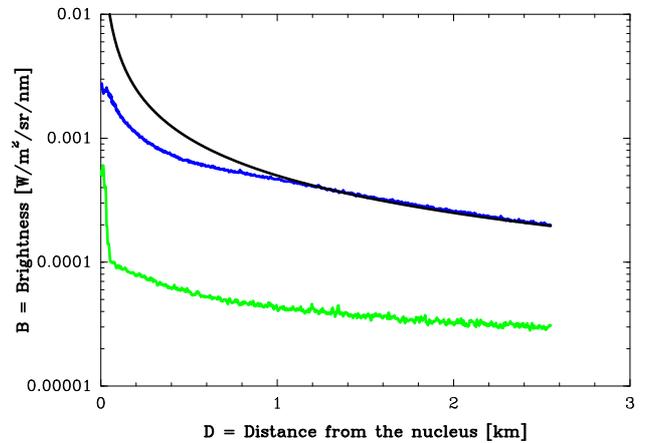}
  \vspace{0.3in}
\caption{2015-07-29T13:25:28 - In blue is the radial profile for the outburst and in green is the radial profile for the coma background. In black is the radial profile over a cone. }
\label{Method_Plot}
\end{figure}

Figure \ref{Method_Plot} shows the radial brightness of the outburst  (after subtraction of the background coma) and the background coma. In comparaison, we added the dispersion of the gas and dust as a function of the distance from the nucleus. As explained by \cite{Gicquel_2016}, we assume $B$ $\propto$ $D^{\beta}$, where $B$ is the brightness, $D$ is the radial distance from the surface of the nucleus and $\beta$ is the slope of log$B$ vs. log$D$. For $D$ $>$ 1km, the brightness profile of the outburst, $\beta$ = 0.94, is much steeper than the brightness profile of the background coma, $\beta$ = 0.41. The outburst seems to follow a divergent pattern for a distance from the nucleus of $D$ $>$ 1km. However, we can see a bump in the radial profile of the outburst and the coma background at $D$ $\approx$ 50 m. Consequently, we anticipated that the outburst was a combination of gas and dust.

\section{Model}
We used the Direct Simulation Monte Carlo (DSMC) method implemented in PI-DSMC (www.pi-dsmc.com) to study the gas flow close to the nucleus and the dust trajectories. The DSMC method is typically the method of choice to study the gas flow in the coma due to its applicability over a large range of Knudsen numbers. Our model produces artificial images for a wide range of parameters, including the gas production rate at the surface, the surface temperature, and the properties of the dust grains. In detail, the model uses the velocity field and the density field obtained with the DSMC to compute the drag force acting on the moving dust particles. 
The drag force $F_{drag}$ is defined as :
\begin{equation}
F_{drag}(r) = \frac{1}{2}(v_{gas}(r)-v_{particle})^2 \rho_{r} \sigma_{CS} C_D
\end{equation}
where $v_{gas}$ is the gas velocity along the radial distance from the nucleus $r$, $v_{particle}$ is the grain velocity, $\rho_{r}$ is the gas density and $\sigma_{CS}$ is the particule cross section and $C_D$ is the drag coefficient of grains. Trajectories are obtained by integration of the equation of motion that also contains the gravitational force around the nucleus taking into account the complex shape. The comet is modeled as two masses with a bulk density of the nucleus 532 $\pm$ 7 kg $\rm m^{-3}$ \citep{Jorda_2016}. The mass of the small lobe and the big lobe are 2.7  $\times$ $10^{12}$ kg and 6.6  $\times$ $10^{12}$ kg, respectively. The contribution of a single trajectory to the dust density field is obtained by computing the time a dust particle spends in a volume cell. The final dust field is computed from trajectories of particles starting at selected surface triangles. The final image is obtained by integrating the density of the dust field in columns parallel to the line of sight. In the case of an optically thin environment, the intensity in the image is assumed to be proportional to the integrated density.\\

\begin{figure}
  \centering
  \subfloat[Blackbody Temperature (K)]{\label{TBB_Fougere}\includegraphics[scale=0.25]{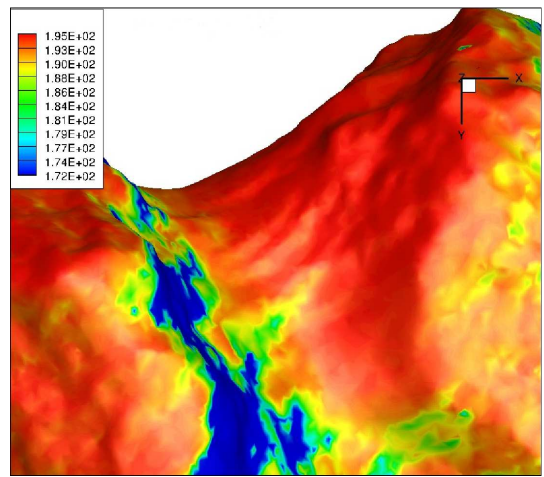}} 
  
  \subfloat[Water production rate ($\rm s^{-1}$$\rm m^{-2}$ ) ]{\label{Qwater_Fougere}\includegraphics[scale=0.25]{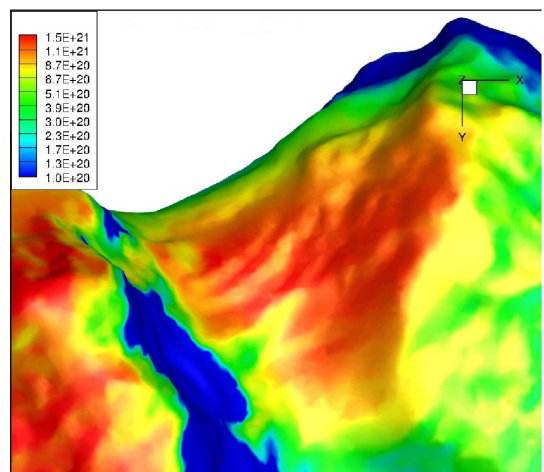}} 
\caption{The blackbody temperature and the water production rate at the surface of the comet \citep{Fougere_2016}}
  \label{Imput_DSMC}
\end{figure}

We used the Direct Simulation Monte Carlo (DSMC) method implemented in PI-DSMC to study the outburst on July 29, 2015. The outgassing rate and the temperature at the surface, from the model described in \cite{Fougere_2016} are shown in Figure \ref{Imput_DSMC}. We assumed a temperature at the surface of $T_{surf}$ = 190 K (Figure \ref{TBB_Fougere}) and a water production rate at the surface of $Q_{H_2O}$ = 3 $\times$ $10^{-5}$ kg $\rm s^{-1}$ $\rm m^{-2}$  (Figure \ref{Qwater_Fougere}). Then, we defined an active region on the surface of 67P at the source location of the outburst. In the case of the active region, we assumed a gas production rate of $Q_{active}$ = $\alpha$ $Q_{H_2O}$, an outgassing ratio between the outburst source and the local region of either 10 or 100, and a temperature of $T_{active}$  = 230 K. Under this model, the change in temperature had no effect on the opening of the outburst. The topography is also taken into account in the model, as \cite{Hoefner_2016} has shown that fractures can be a heat trap, within specific illumination conditions.\\

The simulation uses a Cartesian mesh from which the collision cells and the sampling cells are built up. The collisions between gas molecules are computed using the hard sphere model \citep{Bird_1994}. The colliding molecules are the nearest neighbors, and the size of the simulated domain is 600  $\times$  600  $\times$  1.100 m. In the case of $\alpha$ = 10, the number of collision cells is 21,096,584 and the size of each individual cell is 2.42 m. In the case of $\alpha$=100, the number of collision cells is 10,481,915 and the size of each individual cell is 3.05 m. Also, particles hitting the surface are reflected with a velocity distribution corresponding to the surface temperature.\\

 \section{Results}
\label{Sec_Results}

\begin{figure}
  \centering
  \subfloat[Source location of the outburst in red]{\label{ShapeView}\includegraphics[scale=0.41]{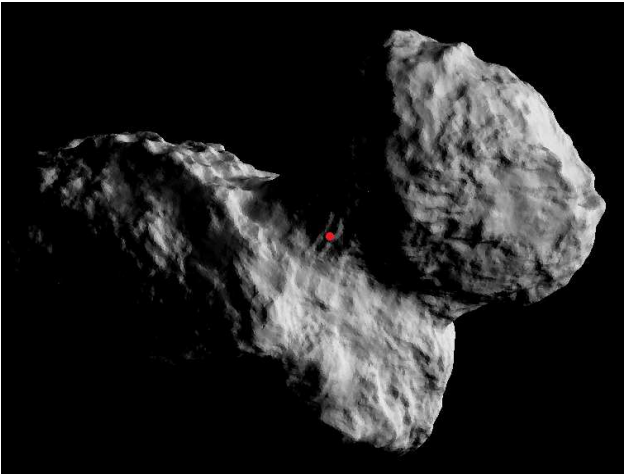}} 
  
  \subfloat[Region around the outburst in black]{\label{snapshot2}\includegraphics[scale=0.22]{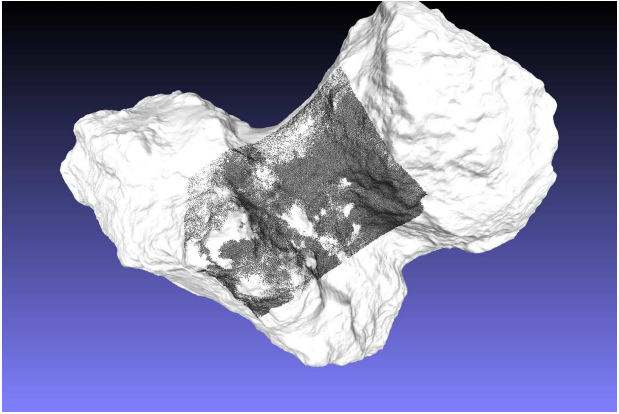}} 
  
  \subfloat[Number density on surface of the nucleus ($m^{-3}$)]{\label{Nucleus}\includegraphics[scale=0.185]{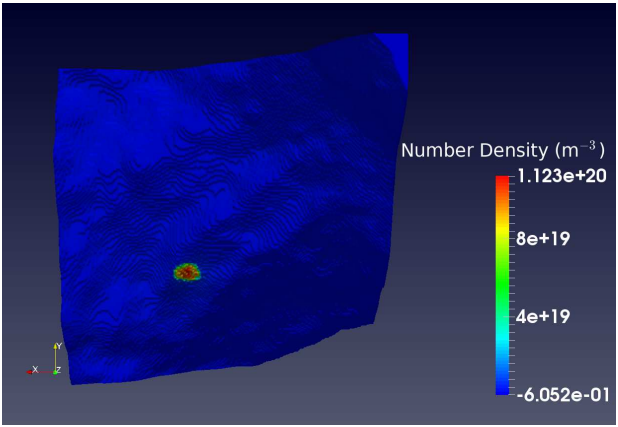}} 

\caption{The method and results for the DSMC model}
  \label{Image_NAC}
\end{figure}

Our model was used to simulate the mechanisms that produced the outburst of July 29, 2015. The source location of the outburst is shown in Figure \ref{ShapeView}. Using the shape model shap5-v1.5-cheops-800k developed by \cite{Jorda_2016}, we examined a region around the outburst, as shown in Figure \ref{snapshot2}. The surface temperature and water production rate at the surface of the nucleus is given in Figure \ref{Imput_DSMC}. We created an active surface with a higher gas production rate at the localization of the outburst, shown in Figure \ref{Nucleus}. The model, as described in Section 2, produced a  series of synthetic images, and we then compared them with the OSIRIS observations.  \\

For purposes of this paper, we assumed that the outburst is composed of only gas (water) and dust. Because the dust is brighter than the gas, the OSIRIS cameras captured brighter images of the dust. In order to simulate the entire outburst, we needed to first simulate only the gas. We then incorporated the dust into the same model used to create the simulated images. \cite{Combi_2012} explained that the gas and the dust have very different behavior, notably regarding their expansion when they are released from an active area. Dust particles receive most of their acceleration by the gas just above the small active area and are accelerated to much larger terminal velocities.  \\
 
\begin{figure}
  \centering
  \subfloat[Velocity]{\label{10x_Velocity}\includegraphics[scale=0.26]{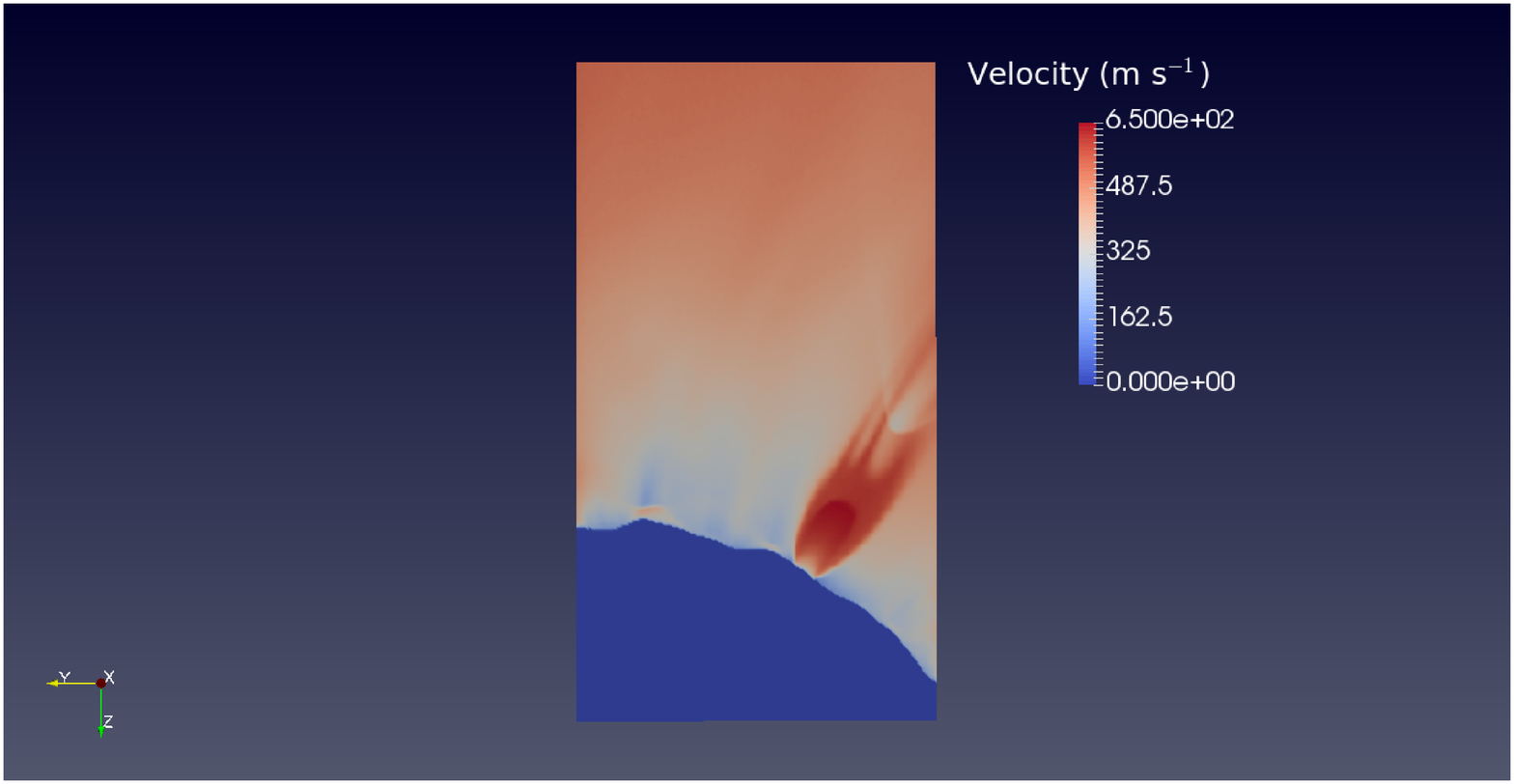}} 
  
  \subfloat[Number density ($m^{-3}$)]{\label{10x_Nb_Density}\includegraphics[scale=0.26]{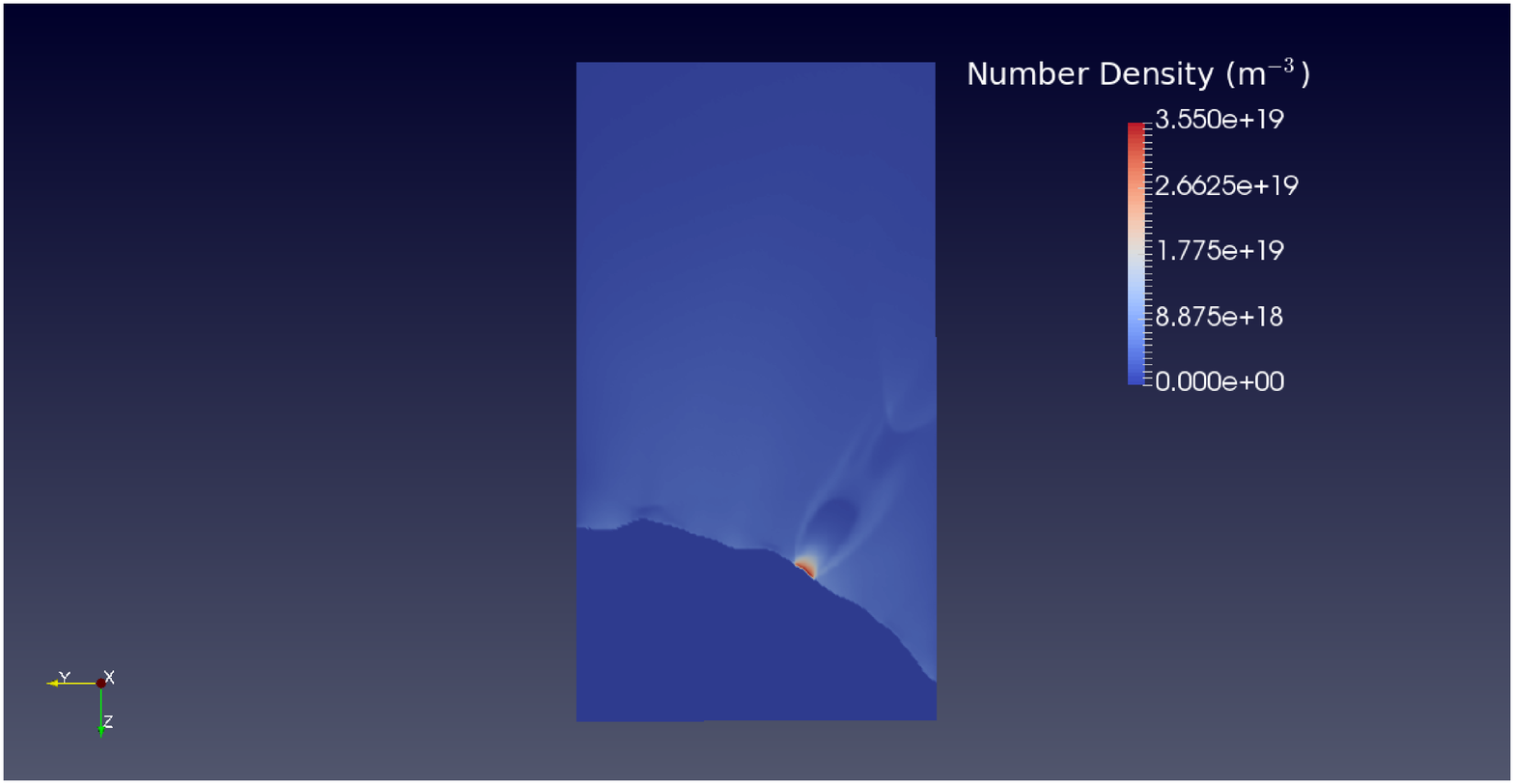}} 
  
  \subfloat[Column density ($m^{-2}$)]{\label{10x_Col_Density}\includegraphics[scale=0.23, angle=330]{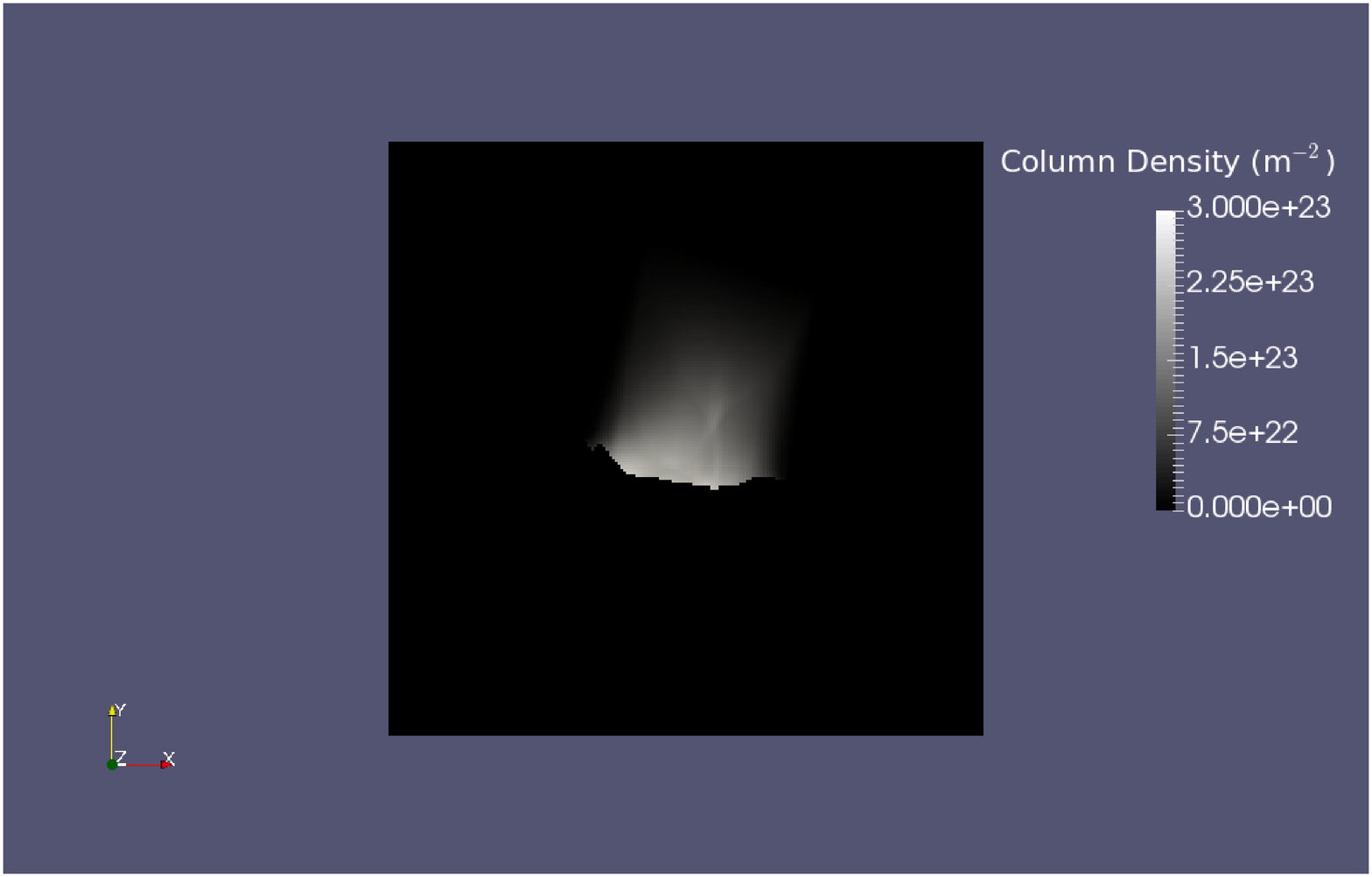}} 

\caption{The results of the DSMC model for water gas at $\alpha$ =10}
  \label{Number_Density_10}
\end{figure}

\begin{figure}
  \centering
  \subfloat[Velocity]{\label{100x_Velocity}\includegraphics[scale=0.26]{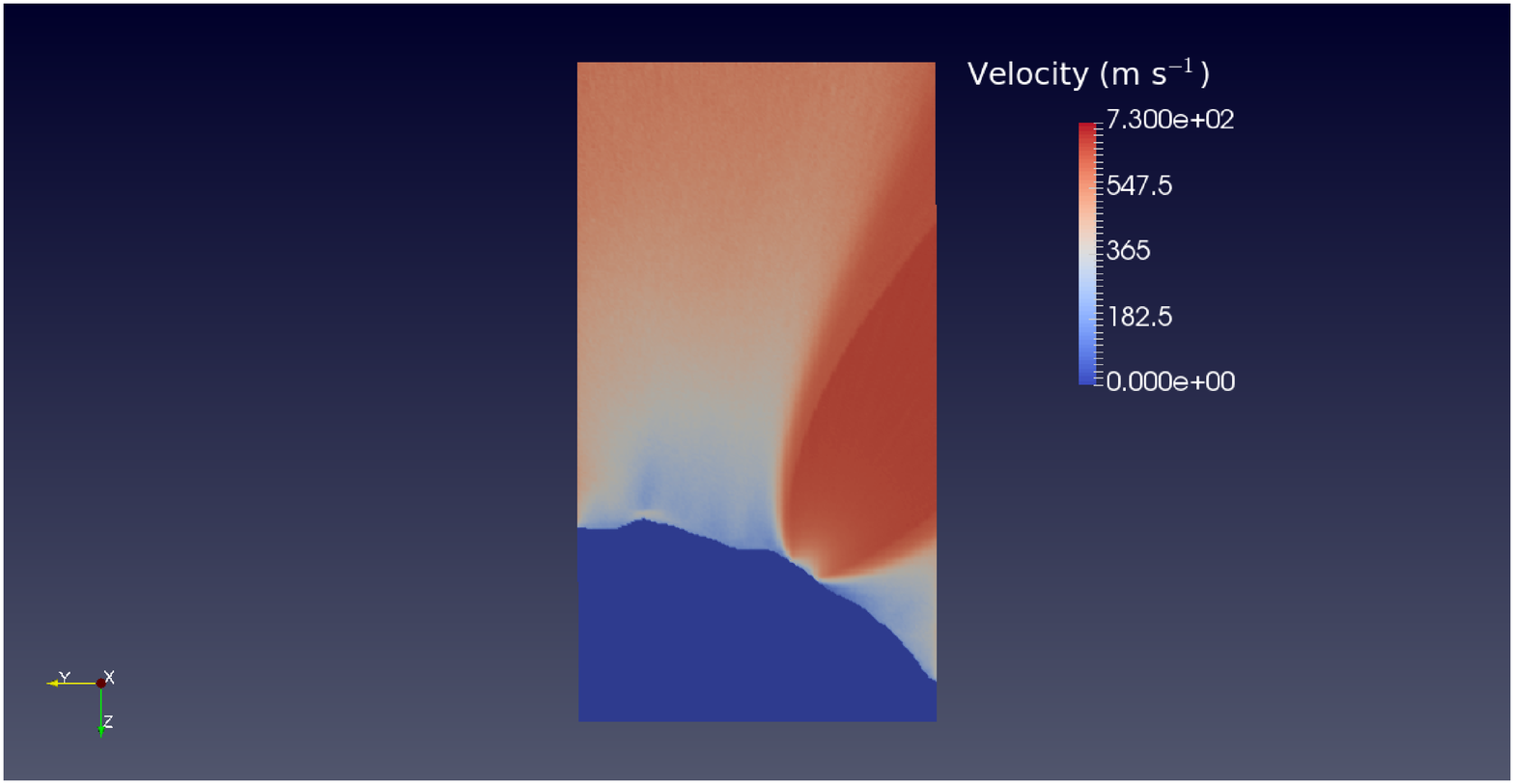}} 
  
  \subfloat[Number density ($m^{-3}$)]{\label{100x_Nb_Density}\includegraphics[scale=0.26]{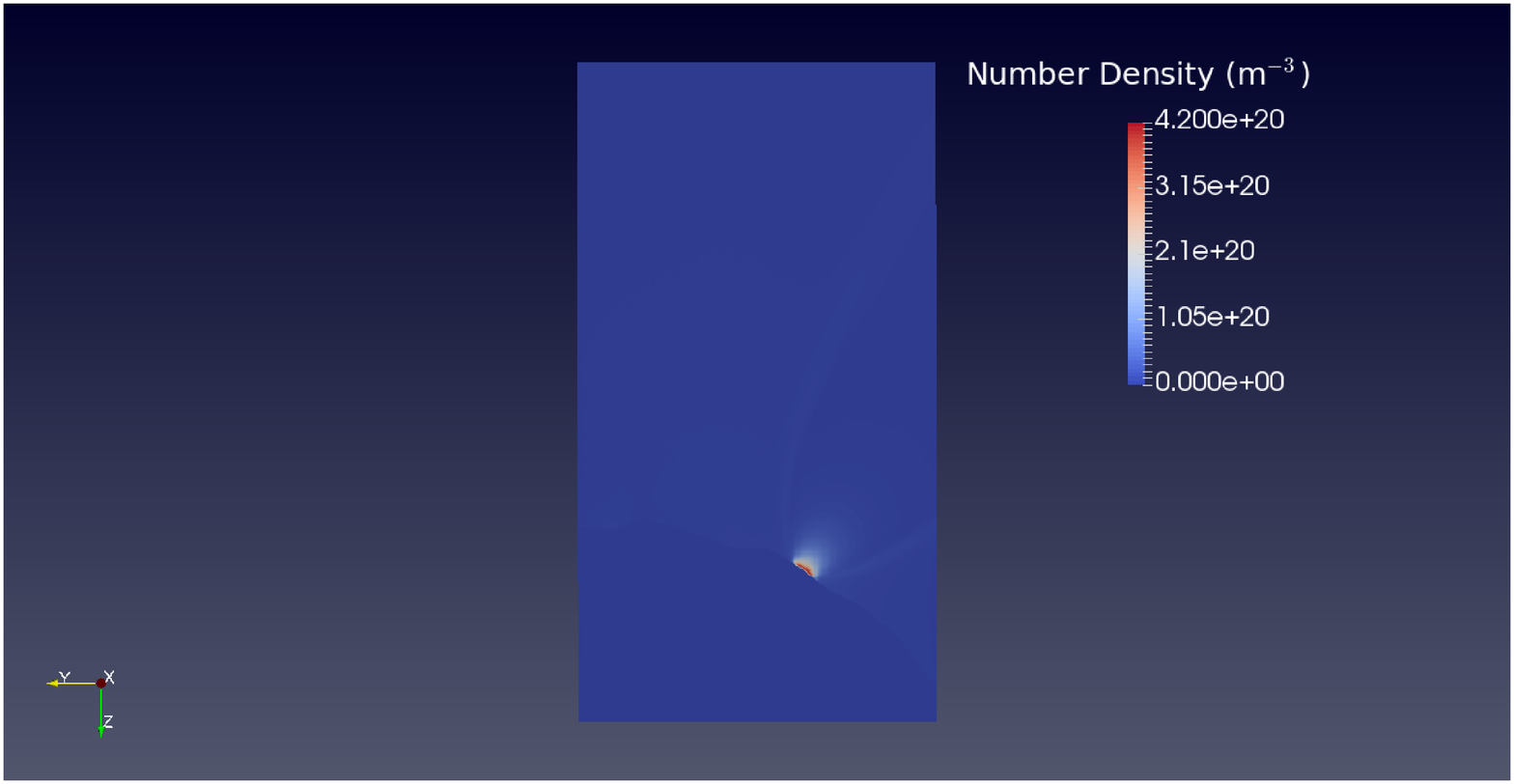}} 
  
  \subfloat[Column density ($m^{-2}$)]{\label{100x_Col_Density}\includegraphics[scale=0.23, angle=330]{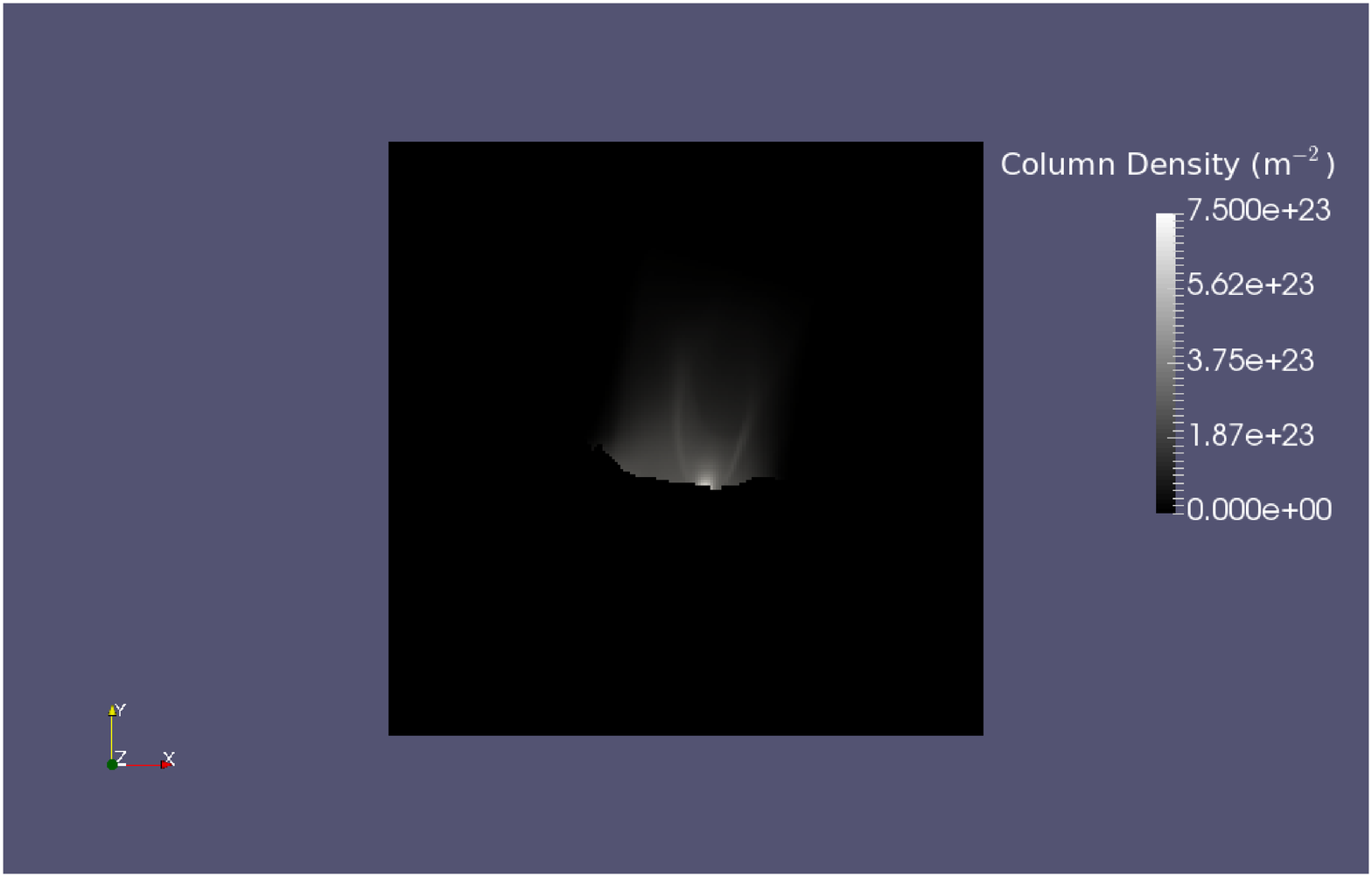}} 

\caption{The results of the DSMC model for water gas at $\alpha$ =100}
  \label{Number_Density_100}
\end{figure}

Throughout Figures \ref{Number_Density_10} and \ref{Number_Density_100}, we used the velocity and number density of only the gas to verify the point of convergence in the gas field.  The size (315 x 585 pixels) and the position in the WAC FOV of the images from the simulation are shown in Fig \ref{VIRTIS_Model} (green box). As shown in  the corresponding Figures \ref{10x_Velocity}, \ref{10x_Nb_Density}, \ref{100x_Velocity} and \ref{100x_Nb_Density}, we plotted the velocity and the number density in the Y-Z plane. The coordinate system that we used in the model was aligned with the coordinate system from the shape model. In the case of $\alpha$ = 10 (Figure \ref{Number_Density_10}) and $\alpha$ = 100 (Figure \ref{Number_Density_100}), the maximum outflow velocity was 650 m s$^{-1}$ and 730 m s$^{-1}$, respectively. The number density reached a maximum around 3.6 $\times$ 10$^{19}$ m$^{-3}$ and 4.2 $\times$ 10$^{20}$ m$^{-3}$ for $\alpha$ = 10 and $\alpha$ = 100, respectively. We then integrated the number density along the line of sight to derive the column density, which is shown in Figures \ref{10x_Col_Density} and \ref{100x_Col_Density}. The high column density close to the nucleus can explain the bump seen in the radial profile  D $\approx$ 50 m (Figure \ref{Method_Plot}). \\

\begin{figure}
  \centering
  \subfloat[Trajectories of the dust (radius 1.97 $\mu$m)]{\label{Number_Density_10_dust-7}\includegraphics[scale=0.55, angle=330]{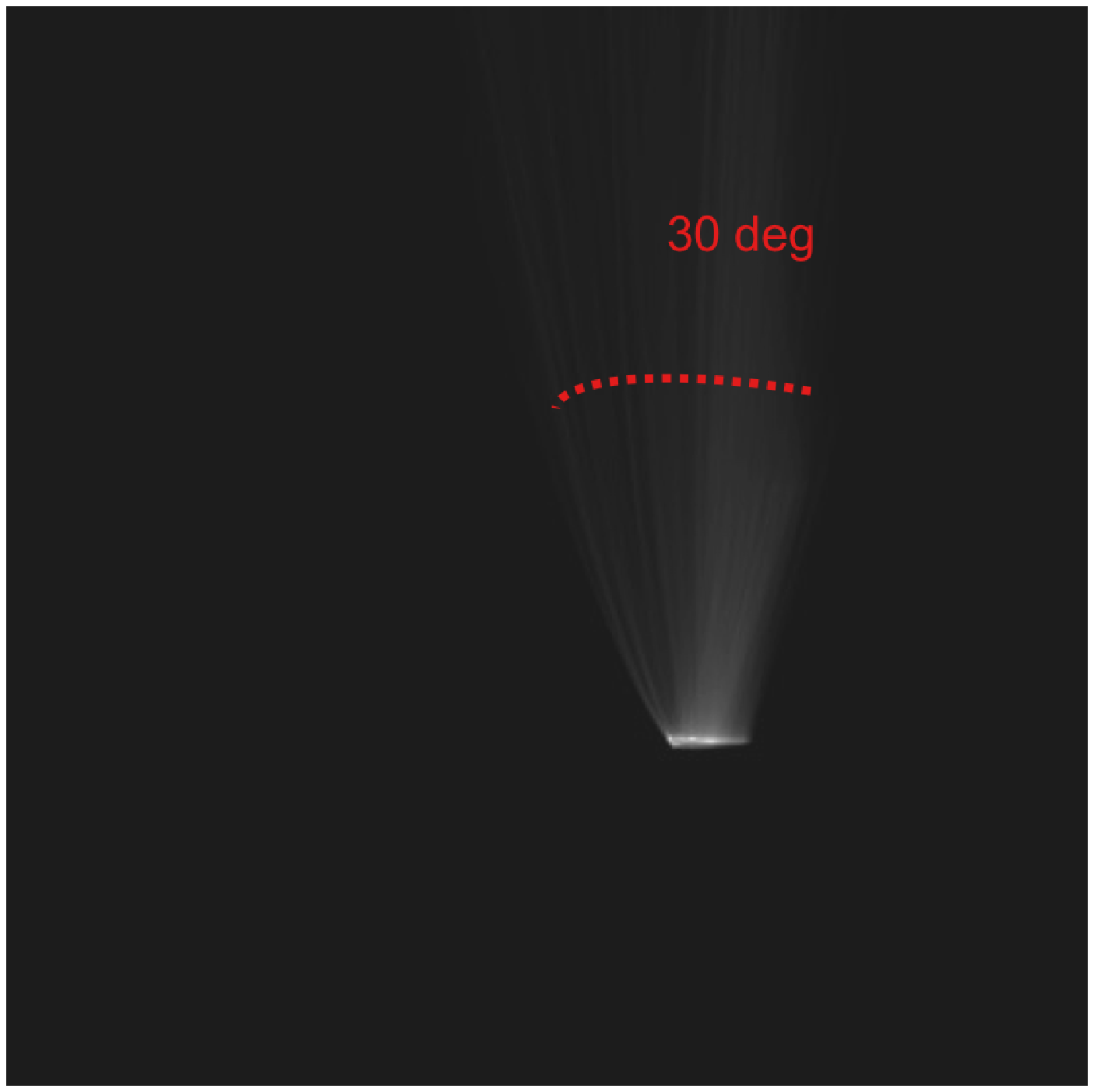}} 
  
  \subfloat[Trajectories of the dust (radius 185 $\mu$m)]{\label{Number_Density_10_dust-13}\includegraphics[scale=0.55, angle=330]{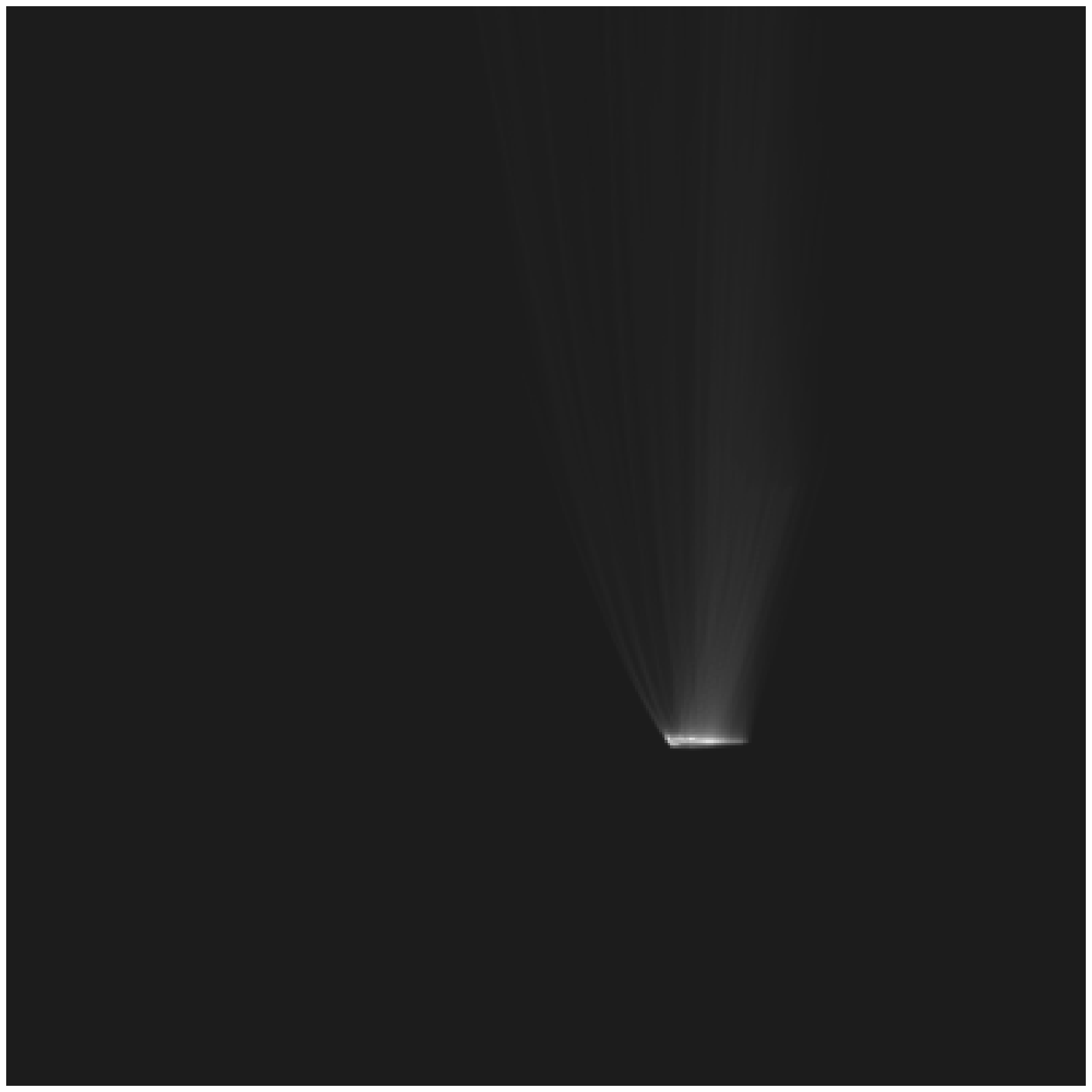}} 

\caption{The results of the DSMC model for the dust at $\alpha$ =10}
  \label{Number_Density_10_dust}
\end{figure}

\begin{figure}
  \centering
  \subfloat[Trajectories of the dust (radius 01.97 $\mu$m)]{\label{Number_Density_100_dust-7}\includegraphics[scale=0.55, angle=330]{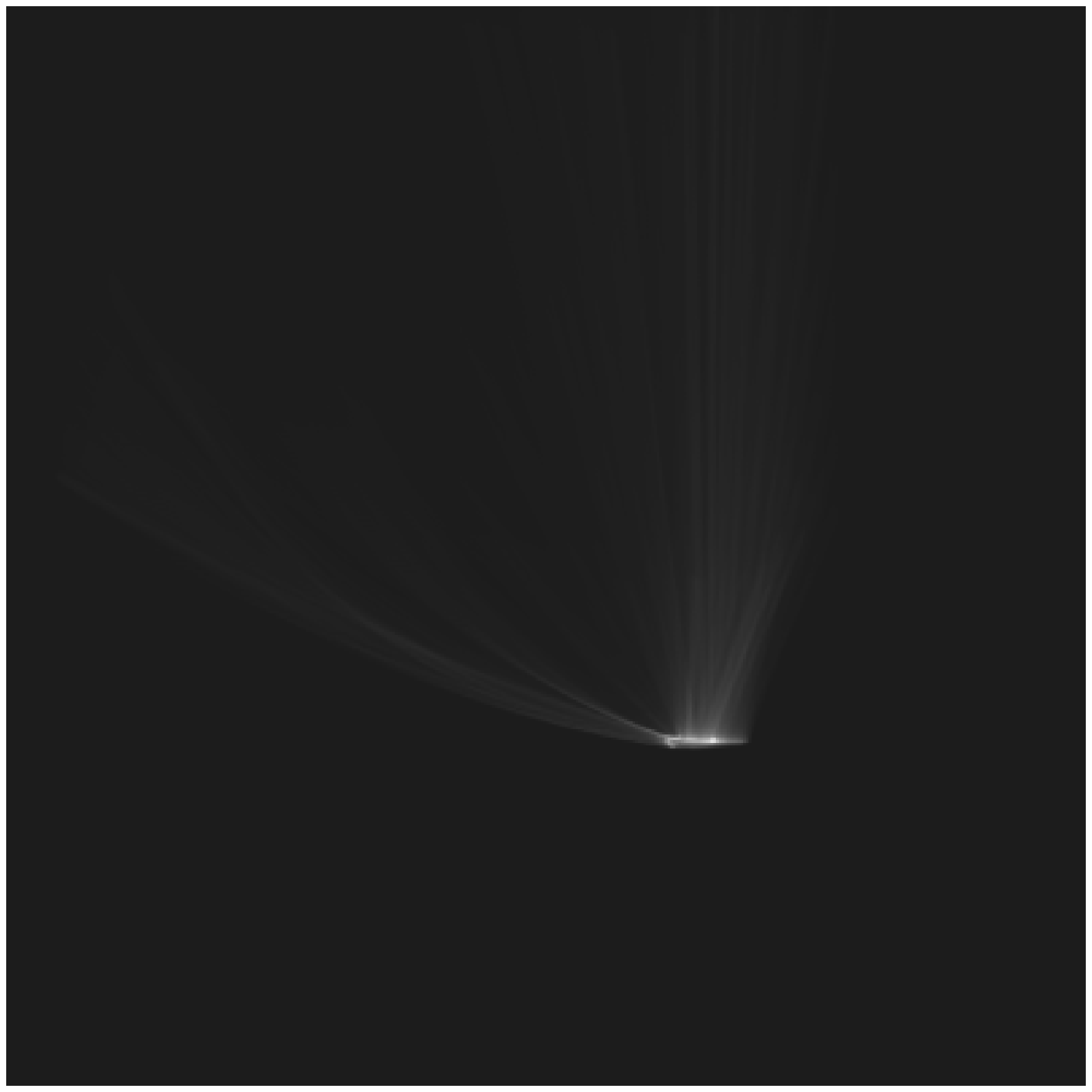}} 
  
  \subfloat[Trajectories of the dust (radius 185 $\mu$m)]{\label{Number_Density_100_dust-13}\includegraphics[scale=0.55, angle=330]{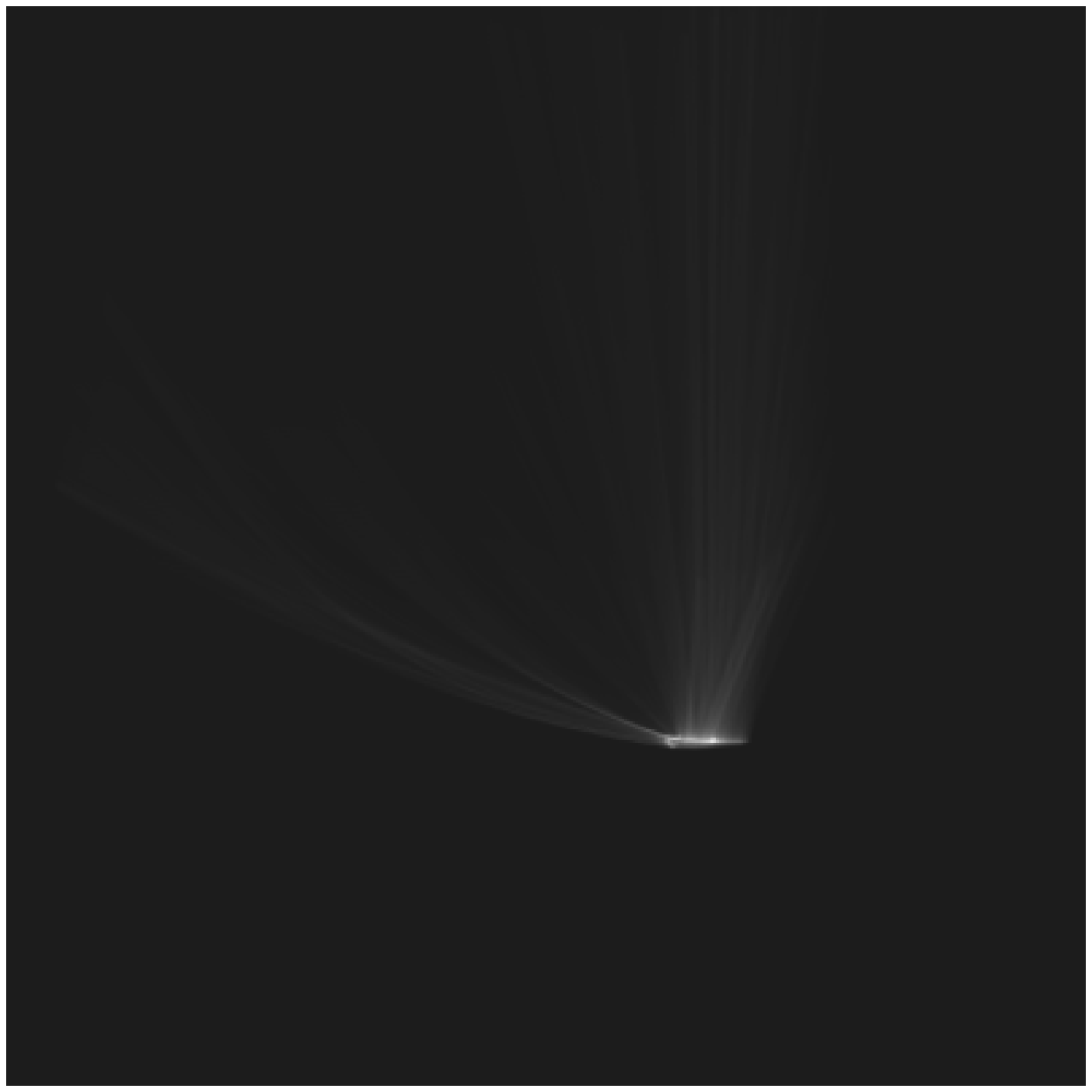}} 

\caption{The results of the DSMC model for the dust at $\alpha$ =100}
  \label{Number_Density_100_dust}
\end{figure}

The results of the simulations that incorporated the dust are shown in Figures \ref{Number_Density_10_dust} and \ref{Number_Density_100_dust}. We know that there are multiple contributions to the brightness, for example: the sun light is scattered by the dust and the light is generated by physical and chemical processes occurring in the gas. The dust was introduced in the simulation to model the light scattered by the dust particles. This included the region close to the nucleus but also the region far away from the nucleus. The brightness in the image corresponded to the column density of dust particles. The assumption was that each dust particle scatters light from the sun into the camera. The intensity in the image was assumed to be proportional to the integrated dust density. In this particular study, the radius of the dust particles are 1.97 $\mu$m (Figures \ref{Number_Density_10_dust-7} and \ref{Number_Density_100_dust-7}) and 185 $\mu$m (Figure \ref{Number_Density_10_dust-13} and \ref{Number_Density_100_dust-13}) according to \cite{Muller_1999}. This is in the size range obtained by \citep{Grun_2016} and by \cite{Lin_2017}. In the case of this model, the synthetic images show little dependence on the particle size. The simulations that included the dust produced images that were even more similar to the actual images obtained with the NAC camera. In Figure \ref{Number_Density_10_dust} the active surface was set at a gas production rate 10 times higher than the base rate for the other parts of the surface of the nucleus. In this case, the dust was even more collimated; the opening angle was within 30 degrees; and the dust projected further out from the surface of the comet. This shape and opening angle correspond to the images obtained by the NAC camera on July 29, 2015. In Figure \ref{Number_Density_100_dust}, we set the gas production rate at 100 times the base rate. At this rate, the model did not reproduce the shape of the outburst; instead, the opening angle on the dust is much wider.\\

At this wavelength, the NAC is more sensitive to the dust. As a result, we concluded that the outburst was in fact a combination of both gas and dust, in which the active surface was generating dust at a gas production of approximately 10 times higher than the base rate found at the nucleus.  \\

The comparison between the model and the OSIRIS image gives us an indication of the number of dust particles ($N_{dust}$) we need to reproduce the observed brightness flux, $B$, in the OSIRIS image. The theoretical brightness for a dust particle $I$ (W $\rm m^{-2}$ $\rm nm^{-1}$ $\rm sr^{-1}$) is given as:

\begin{equation}
	I=\frac{A\phi\left(\alpha\right)}{\pi} \frac{F_{Sun,\lambda_{VIS}}}{R_h^2}\frac{1}{\Delta_{S/C}^2}  \pi a^2 \frac{1}{A_{px}}
\end{equation}
where $A$ = 6.5 $\times$ $10^{-2}$ is the geometric albedo; $\alpha$ = 90 deg is the phase angle; $\phi(90)$ = 0.02 is the phase function \citep{Fornasier_2015}; $F_{Sun,\lambda_{ORANGE}}$ = 1.5650 W $\rm m^{-2}$ $\rm nm^{-1}$ is the flux of the Sun at the central wavelength of the orange filter; and $A_{px}$ =  3.5 $\times$ $10^{-10}$ steradian is the solid angle of a single pixel.\\

The number of dust particles we need to reproduce the observed brightness flux in Figure \ref{Method_Plot} is: $N_{dust}$ = $B$$\times$$L_{px}$/$I$, where $L_{px}$ = 1000 px is the length of the outburst. The total mass of dust (kg) is given by: $M_{dust}$ = (4/3) $\pi$ $a^3$ $\rho$ $N_{dust}$, where $\rho$ = 1000 kg $\rm m^{-3}$ is the bulk density \citep{Grun_2016}. To reproduce the data we need 7.83 $\times$ $10^{11}$ < $N_{dust}$ <  6.90 $\times$ $10^{15}$, for 1.97 $\mu$m < $a$ < 185 $\mu$m. The total mass of dust particles correspond to 220 kg < $M_{dust}$ < 21 Tonnes. This number is in good agreement with the mass estimated by \cite{Vincent_2016}, with \cite{Grun_2016} and \cite{Lin_2017}.

\section{Discussion and conclusion}

The mechanisms that produce the outburts observed on bodies throughout the solar system are still not fully understood.  For this study, we examined one outburst out of many from a group known as the 'summer fireworks', which were observed on the surface of comet 67P/Churyumov-Gerasimenko around the perihelion \citep{Vincent_2016}.\\

We reviewed a number of images taken on July 29, 2015 by the OSIRIS NAC camera in order to precisely determine the source of this outburst on the surface of the comet. The outburst location was in the Sobek region, at a latitude = -37$^{\circ}$ and longitude = 300$^{\circ}$ \citep{Vincent_2016}.  As a number of mechanisms including the morphology of the surface of the comet were likely responsible for the production of the outburst, we decided to use a shape model including the topography. In this particular case, the localization of the outburst was between two hills \citep{Vincent_2016b}. \cite{Skorov_2016} developed a model to explain the outbursts from fractured terrains based on the thermophysics, morphology and composition of the surface. They concluded that close to perihelion, the stresses on the nucleus led to a release of gas and dust. Additionally, the sublimation of icy grains on the surface almost certainly plays a role. Because of the insolation, the temperature increases, possibly creating the jet \citep{Gicquel_2016, Keller_2015, Lin_2016}.  \\

Using the DSMC method, we generated a number of artificial images that attempt to recreate the outburst seen on July 29, 2015 with a gas production rate at the source point of the outburst about 10 times the background production. When accounting only for the gas flow, we were not able to reproduce the observed outburst. It was not until the dust field was integrated into the model that we were able to simulate images that approximate the shape and angle of the outburst, including a noticeable bump in the radial profile at D $\approx$ 50 m. To reproduce the data we need a number of dust particles  7.83 $\times$ $10^{11}$ -  6.90 $\times$ $10^{15}$ (radius 1.97 - 185 $\mu$m), which corresponds to a mass of dust 220 - 21 $\times$ $10^{3}$kg. \\

This is the first publication using this specific model and technique. The ability to successfully reproduce the opening angle and the overall shape of the outburst is useful. More significant is the ability to simulate the potential role of both the gas and the dust in the formation of an observed outburst.  Future simulations using this model and other models can better our understanding of observed events. In the future, we should compare these initial results to future simulations to answer several basic questions: What models best reproduce the observed event? What differences if any exist?  What other assumptions can be made?  This technique can have broad applicability not only to outbursts on comets but also potentially similar phenomenon observed on icy bodies in the solar system.  Well formulated assumptions are critical to our understanding of observed events; however, it is also important to develop new techniques and tools to test our assumptions.  In this paper, we can provide a an estimate for the mass of the ejected dust and for the first time explain the mechanisms producing a single outburst by comparing a model with observation.\\

\section*{Acknowledgements}

OSIRIS was built by a consortium of the Max-Planck-
Institut f{\"u}r Sonnensystemforschung, G{\"o}ttingen, Germany, CISAS University of Padova, Italy, the Laboratoire d'Astrophysique de Marseille, France, the
Institutode Astrof\'isica de Andalucia, CSIC, Granada, Spain,the Research and
Scientific Support Department of the European Space Agency, Noordwijk, The
Netherlands, the Instituto Nacionalde T\'ecnica Aeroespacial, Madrid, Spain, the
Universidad Polit\'echnica de Madrid, Spain, the Department of Physics and
Astronomy of Uppsala University, Sweden, and the Institut f{\"u}r Datentechnik und
Kommunikationsnetze der Technischen Universit{\"a}t Braunschweig, Germany.
The support of the national funding agencies of Germany (DLR), France(CNES),
Italy(ASI), Spain(MEC), Sweden(SNSB), and the ESA Technical Directorate
is gratefully acknowledged. We thank the Rosetta Science Ground Segment at
ESAC, the Rosetta Mission Operations Centre at ESOC, and the Rosetta Project
at ESTEC for their outstanding work enabling the science return of the Rosetta
Mission.

The research was carried out at the Jet Propulsion Laboratory, California Institute of Technology, under a contract with the National Aeronautics and Space Administration.


\bsp	
\label{lastpage}
\end{document}